\documentclass[onecolumn, letterpaper, 10 pt, doublespace]{IEEEtran}  

\IEEEoverridecommandlockouts                              

\expandafter\def\expandafter\normalsize\expandafter{%
    \normalsize
    \setlength\abovedisplayskip{3pt}
    \setlength\belowdisplayskip{3pt}
    \setlength\abovedisplayshortskip{3pt}
    \setlength\belowdisplayshortskip{3pt}
}

\usepackage{graphicx}
\usepackage[perpage]{footmisc}
\usepackage{caption}
\usepackage{tabularx}
\usepackage{color}
\usepackage{wrapfig}
\usepackage{epsfig} 
\usepackage{mathptmx} 
\usepackage{times} 
\usepackage{amsmath} 
\usepackage{amssymb}  
\usepackage{cite} 
\usepackage[space]{grffile}
\usepackage{epstopdf}
\usepackage{pdfpages}
\usepackage[mathscr]{euscript}
\usepackage[utf8]{inputenc}
\usepackage{lipsum}
\usepackage[normalem]{ulem} 
\usepackage{algorithm,algorithmic}
\DeclareMathOperator*{\argmin}{arg\,min}

\title{Sub-optimal Control of Autonomous Wheel loader with Approximate Dynamic Programming  }

\author{Tohid Sardarmehni$^{1}$ and Xingyong Song $^{2,*}$
        \thanks{$^{1}$ Tohid Sardarmehni, Postdoctoral Research Fellow,	Department of Engineering Technology and Industrial Distribution, Texas A\&M University, College Station, TX, 77843, USA.
{\tt tsardarmehni@smu.edu}}%

\thanks{$^{2,*}$ Xingyong Song, Assistant Professor, Department of Engineering Technology and Industrial Distribution; Department of Mechanical Engineering; College of Engineering; 	Texas A\&M University, College Station, TX, 77843, USA (Corresponding Author).
	  {\tt songxy@tamu.edu}
    }
\thanks{This research was partially supported by the National Science Foundation under Grant No. 1826410.  }
\thanks{This article has been accepted for oral presentation at 2019 Dynamic System and Control Conference. The content might NOT be the same as the final edition of the accepted paper for the DSCC.  }
}
\usepackage{graphicx}
\usepackage{tabularx}
\usepackage{color}
\usepackage{wrapfig}
\usepackage{epsfig} 
\usepackage{mathptmx} 
\usepackage{times} 
\usepackage{amsmath} 
\usepackage{amssymb}  
\usepackage{cite} 
\usepackage[space]{grffile}
\usepackage{epstopdf}
\usepackage[mathscr]{euscript}
\usepackage[utf8]{inputenc}
\usepackage{lipsum}
\usepackage{mathtools}
\usepackage[margin=0.5cm]{caption}
\usepackage{graphicx}
\usepackage{caption}
\usepackage{tabularx}
\usepackage{wrapfig}
\usepackage{epsfig} 
\usepackage{mathptmx} 
\usepackage{times} 
\usepackage{amsmath} 
\usepackage{amssymb}  
\usepackage{cite} 
\usepackage[space]{grffile}
\usepackage{epstopdf}
\usepackage[utf8]{inputenc}
\usepackage{color}
\usepackage{slashbox}
\usepackage{lipsum}
\usepackage{float}
\usepackage{placeins}
\usepackage[normalem]{ulem} 
\usepackage{algorithm}
\usepackage{algorithmic}
\usepackage[algo2e]{algorithm2e} 
\usepackage{setspace}
\usepackage{mathrsfs}

\DeclareMathOperator{\diag}{diag}

\newtheorem{Rem}{Remark}

\begin{document}
\maketitle   

\begin{abstract} \label{Abstract}
Optimal control of wheel loaders in short loading cycles is studied in this paper. For modeling the wheel loader, the data from a validated diesel engine model is used to find a control oriented mean value engine model. The driveline is modeled as a switched system with three constant gear ratios (modes) of $-60$ for backwarding, $60$ for forwarding, and zero for stopping. With these three modes, the sequence of active modes in a short loading cycle is fixed as backwarding, stopping, forwarding, and stopping. For the control part, it is assumed that the optimal path is known a priori. Given the mode sequence, the control objective is finding the optimal switching time instants between the modes while the wheel loader tracks the optimal path. To solve the optimal control problem, approximate dynamic programming is used. Simulation results are provided to show the effectiveness of the solution.
\end{abstract}


\begin{table*}[htbp]
\caption{Nomenclature}
\label{table_notations}
\centering
\begin{tabular}{ | c | l | c | l | }
\hline
		$\omega_e$ & rotational speed of engine  &  $\alpha$ & angle between lift force and lift arm \\ \hline
    $J_e$ & engine moment of inertia  & $\theta$ & lift arm angle\\ \hline
    $T_e$ & engine torque  & $T_{buc}$ & torque due to the bucket load  \\ \hline
		$p_{im}$ & intake manifold pressure & $T_{arm,w}$ & torque due to boom weight  \\ \hline
		$u_f$ & mass of fuel per engine cycle  & $I_{boom}$ & boom moment of inertia  \\ \hline
		$\omega$ & angular velocity of the boom  & $F_w$ & wheels' traction force  \\ \hline
		$V$ & vehicle velocity & $F_{roll}$ & rolling resistance force  \\ \hline
		$u_p$ &pressure in the lift cylinder  & $M_{tot}$ & mass of WL and load  \\ \hline
		$u_s$ & time derivative of steering angle  &  $X$ & position of WL\\ \hline
		$\tau_p$ & time constant in $p_{im}$ & $Y$ & position of WL  \\ \hline
		$p_{stat}$ & stationary intake manifold pressure  &  $\beta$ & heading angle   \\ \hline
		$F_{cyl} $ & lifting force &  $R_{turn}$ & turning radius  \\ \hline
		$R$ & a parameter in the boom geometry &  $\delta$ & steering angle\\ 
		\hline
\end{tabular}
\end{table*}

\section{Introduction} \label{sec_intro}

Wheel loaders (WLs) are essential equipment in off-road constructions. From the control perspective, minimizing fuel consumption and minimizing the operation time of wheel loaders are two crucial and contradictory control goals. 

In general, in a WL an engine provides the power. The three actions which consume power in a WL are traction, lifting, and steering. Traction is linked to the driveline for moving the vehicle or stopping it. Lifting is related to the hydraulic system to lift or bring down the boom/bucket, and steering is the one that navigates the vehicle.  
 
In a construction field, wheel loaders mostly go through repetitive cycles to pick up a load from one point and drop it in another point. This cycle includes 4 actions as backwarding, stopping, forwarding and then stopping. After filling the bucket with a load, the vehicle accelerates backward from the resting point. At time $t_1$ the operator applies the brakes while putting the gearbox in the neutral mode to bring the vehicle to a full stop at time $t_2$. Then the operator puts the gearbox in the forward mode and accelerates toward the unloading point. At time $t_3$, the operator applies the brakes again while putting the vehicle in the neutral gearbox to bring the vehicle to full stop at time $t_4$. When the vehicle stops, the operator unloads the bucket. The path that a WL follows in such a cycle has a V-shape, and it is called a Short Loading Cycle (SLC). 

There are some critical questions in autonomous control of a WL such as 
\begin{itemize}
\item	[1-] What is the best SLC to follow? 
\item	[2-] What are the optimal times to switch from one mode to another mode? 
\item	[3-] What is the optimal time of operation? 
\item	[4-] When is the optimal time to bring the bucket down or lift it? 
\item	[5-] What is the optimal steering angle in an SLC? 
\item	[6-] How can one get the minimum fuel consumption? 
\end{itemize}

In this paper, we tried to address some of these questions in the context of optimal control and left the rest for our future research. The behavior of a WL in an SLC portrays a switched system with controlled subsystems and a fixed mode sequence \cite{Heydary_fix_mode}. Switched systems are systems comprised of several subsystems/modes and a switching rule that assigns the active mode \cite{ Tohid_TNNLS, Tohid_neurocomputing, zahng_2019}. The sequence of active modes in a switched system is called the mode sequence. For example, consider a system with two modes and only one switching which happens at $t= t_1$. One possible mode sequence is $\{ mode \hspace{3pt}1,\hspace{3pt} mode \hspace{3pt}2 \}$ which means that $\forall t<t_1$, mode 1 is active and for $t\ge t_1$ mode 2 is active. 
%
As seen in this example, when the mode sequence is fixed the controller is only responsible for assigning the switching time instants and not the active mode. Also, when the subsystems include a continuous control, for example steering angle, braking, and hydraulic pressure in a WL, the subsystems are called controlled subsystems. In general optimal control of switched systems is a challenging problem due to discontinuous nature of switching \cite{Tohid_neurocomputing, Tohid_TNNLS, Ata_Heydari_DSCC, Tohid_Heydari}. 

A complete engine model has many variables/parameters, and it is not suitable for control purpose. In order to model a diesel engine, the data from a real engine \cite{ Wahlstrom_Eriksson_2011 } is used to tune the parameters of a mean value engine model, introduced in \cite{ Nezhadali_compare_to_optimal }, which captures only the essential characteristics of the engine, but maintains a good level of accuracy. After the engine is modeled, the driveline of a WL is modeled as a switching system with forwarding, backwarding, and stopping modes. In the present study, it is assumed that the desired path is known. Hence, an optimal control problem is formulated to track the desired path and to find the optimal switching times. 

Compared to the existing literature, the present work neglects the filling and emptying dynamics of the bucket as were studied in \cite{ Sarata_2004 }. Also, the path is assumed to be known a priori. This is unlike \cite{Sarata_2005, Shigeru_2017, Alshaer_2013, Hong_2017, Takei_2015} where the path planning is investigated. The present study borrows most of the dynamical modeling from \cite{ Nezhadali_compare_to_optimal }. Also, the current study follows the same idea for modeling the gearbox as a switched system and finding the optimal switching time that was done in \cite{ Nezhadali_compare_to_optimal }. However, the presented optimal control problem formulation and solutions in this paper are entirely different from the ones discussed in \cite{ Nezhadali_compare_to_optimal }. In \cite{ Nezhadali_compare_to_optimal }, a mixed integer programming solution provided by a numerical solver is used for a specific initial condition. However, the solution developed in this paper is a closed loop feedback optimal policy which means that the solution is valid for all initial conditions in a compact set selected as a domain of interest. Also, since in this paper the mode sequence is enforced, by a change of mode sequence, one can generate the results reported in \cite{ Nezhadali_lift_transport } in a closed loop feedback policy.  Compared to \cite{ OH_2015}, \cite{ Ghabcheloo }, this paper provides the optimal policy in the whole time horizon rather than non-optimal control methods. In \cite{ Frank_2018 } optimal control of a wheel loader for gravel application was performed with dynamic programming. In \cite{ wang_2016 }, an energy management strategy for hybrid wheel loaders was studied with dynamic programming along with an advisory control policy. In general, dynamic programming is a strong method which provides the closed loop feedback optimal policy \cite{ kirk2004optimal }.  However, as the order of the system increases rapid access to memory becomes prohibitive which is known as the curse of dimensionality \cite{ kirk2004optimal }. The method provided in this paper is based on Approximate Dynamic Programming (ADP) which finds the near optimal solution instead of the exact optimal solution. This is in fact a remedy for the curse of dimensionality in dynamic programming. 

The rest of this paper is organized as follows. In section \ref{sec_model}, modeling of the diesel engine and the WL are discussed. In section \ref{sec_optimal_cotnrol}, optimal control problem formulations are presented first, and then the optimal control solution is discussed. Simulation results are provided in section \ref{sec_simulation_resutls}, and section \ref{sec_conclusion} concludes the paper.
 
\section{DYNAMICS OF THE WHEEL LOADER}\label{sec_model}

A model for a wheel loader was introduced in \cite{ Nezhadali_compare_to_optimal } which includes four main parts as powertrain, driveline, steering, and hydraulics. This model can be summarized as follows.

\begin{eqnarray} 
\dot{\omega}_e &=&  \frac{1}{J_e} \big(T_e (\omega_e,p_{im},u_f )- P_{e,load} (\omega_e,\omega,V,u_p,u_s )\big) \label{eq_dot_omega_e}\\
\dot{p}_{im} &=&\frac{1}{\tau_p (\omega_e ) } \big( p_{stat} (\omega_e,u_f )-p_{im} \big)	\label{eq_dot_p_im}\\
\dot{\omega} &=&\frac{F_{cyl}(u_p )  R \sin(\alpha(\theta))-T_{buc}(\theta)-T_{arm,w} (\theta)}{I_{boom} } \label{eq_dot_omega}\\
\dot{\theta}&=&\omega \label{eq_dot_theta}\\
\dot{V}&=&\frac{F_w (\omega_e,V) - sign(V) ( u_b+F_{roll})}{M_{tot}}\label{eq_dot_V}\\
\dot{X}&=&V\cos(\beta)	\label{eq_dot_x}\\
\dot{Y}&=&V\sin(\beta)\label{eq_dot_y}\\
\end{eqnarray}
\begin{eqnarray}
\dot{\beta}&=&\frac{V}{R_{turn}(\delta) }\label{eq_dot_beta}\\
\dot{\delta}&=&u_s\label{eq_dot_delta}
\end{eqnarray}

In the subsequent subsection, engine model and its components are explained as they were needed. Due to page constraints, the interested readers are referred to \cite{ Nezhadali_compare_to_optimal } for the details of the lifting dynamics and the driveline.  

\subsection{Engine Model} \label{sebsec_engine_model}
\begin{figure}[htbp]
\begin{center}
\includegraphics[scale = 0.9]{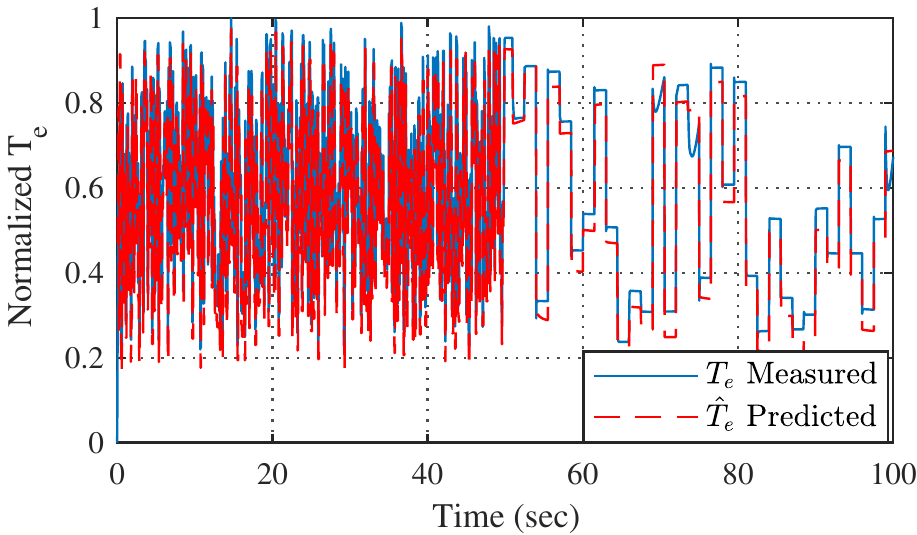}
\end{center}
\vspace{-5pt}	
\caption{Comparison between the predicted $T_e$ and measured $T_e$ with the validation data.} 
\label{Fig_TE_compare}
\vspace{-10pt}
\end{figure}

Equations (\ref{eq_dot_omega_e}) and (\ref{eq_dot_p_im}) are the mean value engine model for modeling the engine torque and manifold pressure. In order to find a function for $T_e$, the data from a validated diesel engine developed in \cite{ Wahlstrom_Eriksson_2011 } is used. Hence, random square wave input signals with short ($0.1$ sec) and long ($1.5$ sec) pulse widths were used to excite both transient and steady-state response of the engine. For gathering the training patterns, the engine model was run for $300$ sec subject to short pulse width and $100$ sec with long pulse width. Then the values of $T_e$, $\omega_e$, $p_{im}$, and $u_f$ were first normalized and then saved. For finding a mean value model, $T_e:\mathbb{R}^3\to\mathbb{R}$ was selected as 
\begin{equation}
\begin{split}
\widehat{T}_e(\omega_e,p_{im},u_f )= W_{T_e}^T [1,\omega_e,p_{im},u_f]^T
\label{eq_predicted_TE}
\end{split}
\end{equation}
where $W_{T_e } \in \mathbb{R}^4$ is a tunable weight vector and superscript $T$ denotes the transpose operator. For finding $W_{T_e}$ least squares in batch mode was used on the entire data. The Mean Absolute Error (MAE) between the predicted engine torque from (\ref{eq_predicted_TE}) and the data measured from the engine for the training was only $7.5\times 10^{-3}$. For validating the model, a new set of random inputs were given to the engine. For gathering the validation data, in the first $50$ sec short pulse width was used and for the second $50$ sec long pulse width was used. The $W_{T_e}$ that was found before was used, without retraining, to predict the engine torque. The MAE between the predicted and measured values of engine torque was only $4 \times10^{-2}$ with the validation data which shows the effectiveness of the training with least squares. The performance of the model introduced in (\ref{eq_predicted_TE}) and the data from the engine in the validation is shown in Figure \ref{Fig_TE_compare}. 
 \vspace{-10pt}

%
 %

In WLs, the engine provides power for hydraulics, traction, and steering. Hence \cite{ Nezhadali_compare_to_optimal }
\begin{equation}
\begin{split}
P_{e,load}=P_{lift} (\omega,u_p )+P_{steering} (u_s)+P_{traction} (\omega_e,V)
\label{eq_p_eload}
\end{split}
\end{equation}
In (\ref{eq_p_eload}), $P_{lift}$ is the power required for lifting and can be defined as 
\begin{equation}
P_{lift} (\omega,u_p )=\max\Big( 0,\frac{Q(\omega) u_p}{\eta_{lift} } \Big)
\label{eq_p_lift}
\end{equation}
In (\ref{eq_p_lift}), $Q(.)$ is the mass flow rate of the hydraulic fluid into the lifting cylinder and $\eta_{lift}$ is the lifting efficiency. 
Also, steering power can be shown as 
\begin{equation}
\begin{split}
P_{steering}=C_{st} u_s^2
\label{eq_P_steering}
\end{split}
\end{equation}
where $C_{st}$ is a constant. For modeling the traction power, \cite{ Nezhadali_compare_to_optimal } proposed a hybrid model based on the gear ratio as
\begin{equation}
\begin{split}
P_{traction} (\omega,V)=P_{pump}(\omega_e,V,\gamma)|sign(\gamma)|
\label{eq_P_traction}
\end{split}
\end{equation}
Assuming three values as $\gamma \in\{-60, 0,+60\}$, one can consider the switching dynamics as backwarding, stopping, and forwarding. 

For identifying the manifold pressure, the same approach as using the data from the validated engine model was used to tune the parameters of the model introduced in \cite{ Nilsson_Turbo_2012 }, \cite{ Rizzoni } as
\begin{equation}
\begin{split}
\tau_p (\omega_e )= \tau_1 \omega_e+\tau_2
\label{eq_tau_p}
\end{split}
\end{equation}
\begin{equation}
\begin{split}
p_{stat}= a_1 \omega_e+a_2 u_f+a_3
\label{eq_p_stat}
\end{split}
\end{equation}
In (\ref{eq_tau_p}) and (\ref{eq_p_stat}), $\tau_1$, $\tau_2$, $a_1$, $a_2$, and $a_3$ are constants. For this tuning, gradient descent algorithm was used with the sequential offline data. Once the training concluded, the model was validated with a new set of data. The MAE between the trained model and the validation data was only $5.3\times10^{-2}$. 

\subsection{State space model}
 
It is desired to show the dynamics of the WL as
\begin{equation}
\begin{split}
\dot{x}(t) &= \bar{f}_v\big(x(t)\big) + \bar{g}_v\big( x(t) \big) u(t), \: v\in = \{1,2,3\}, \: x(0) = x_0 
\label{eq_dynamics_general} 
\end{split}
\end{equation}
where $x\in \mathbb{R}^n$ is the state vector, and Lipschitz functions $\bar{f}_v: \mathbb{R}^n \to \mathbb{R}^n$ and $\bar{g}_v: \mathbb{R}^n \to \mathbb{R}^m$ denote the dynamics of the subsystems. The active mode in time instant $t$ is shown by sub-index $v$, and the set of all subsystems is shown as $\{1,2,3\}$ which corresponds to forwarding, backwarding, and stopping mode in the WL. Equation (\ref{eq_dynamics_general}) shows a switched system dynamics where each subsystem includes a continuous control. In the literature of the switched systems, this type of subsystems is called controlled subsystems. Also, equation (\ref{eq_dynamics_general}) shows that each mode is control affine. 
To show the dynamics of the WL as (\ref{eq_dynamics_general}), let the state vector as 
\begin{equation}
\begin{split}
x(t)=[x_1,x_2,x_3,x_4,x_5,x_6,x_7,x_8,x_9 ]^T 
\label{eq_state_vector}
\end{split}
\end{equation}
where time dependency of $x_{1\le i \le9}$ is dropped for notational simplicity, i.e., $x_i=x_i (t)$. Also, $x_1= \omega_e$, $x_2=p_{im}$, $x_3=\theta$, $x_4=\omega$, $x_5=X$, $x_6=Y$, $x_7= V$, $x_8=\beta$, $x_9=\delta$.  In \cite{ Nezhadali_compare_to_optimal }, the control vector was selected as $u(t)=[u_f,u_p,u_b,u_s]^T$. In this research we are interested to find the control affine dynamics and existence of nonlinear terms with respect to $u_s$ and $u_p$ will not permit such selection for $u(t)$. To solve this problem, similar to the method used in \cite{ Heydari_path}, two new states can be defined as $x_{10}=u_p$ and $x_{11}=u_s$. Taking time derivative from $x_{10}$, and $x_{11}$, one has 
\begin{eqnarray}
\dot{x}_{10} &=& u_1 \label{eq_added_state_1}\\
\dot{x}_{11} &=& u_2 \label{eq_added_state_2}
\end{eqnarray}
Also, one can select $u_3=u_f$  and $u_4=u_b$. Therefore, $u(t)=[u_1,u_2,u_3,u_4 ]^T$.

\begin{Rem} \label{rem_nondim}
In order to normalize the range of variations of the variables, one can nondimensionalize the dynamics. For this purpose, let $X_1 $ as the maximum value for variable $|x_1|$. Hence, one can define $\bar{x}_1 = \frac{x_1}{X_1}$. With this transformation, one can see that $|x_1| \in [0, X_1]$ and $|\bar{x}_1| \in [0,1]$. Considering a scalar system as $\dot{x_1} = f(x_1) + g(x_1) u(t)$, it is straight forward to see $\dot{\bar{x}}_1 = \frac{1}{X_1}\big(f(X_1 \bar{x}_1) + g(X_1 \bar{x}_1) u(t) \big)$. The discussed nondimensionalization is used in this paper for optimal controller design.  
\end{Rem}


\section{OPTIMAL CONTROL FORMUATION} \label{sec_optimal_cotnrol}

Considering the dynamics as (\ref{eq_dynamics_general}) and assuming the mode sequence is known, it is desired to find optimal switching times and a feedback control policy that minimizes the cost function 
\begin{equation}
\begin{split}
J&(x_0 )= \big(x(t_f)- r(t_f)\big)^T S \big( x(t_f) - r(t_f) \big) +  \int_{t_0}^{t_f} \frac{1}{2}\Big(\big(x(t)-r(t)\big)^T \bar{Q} \big(x(t)-r(t)\big)+ u(t)^T \bar{R} u(t) \Big) dt	\label{cost_func_tracking_contin}
\end{split}
\end{equation}
In (\ref{cost_func_tracking_contin}), $t_0 $ is the initial time, $t_f$ is the final time, and $r \in \mathbb{R}^n$ is the reference signal which is a known function of time. $S \in \mathbb{R}^{n\times n}$ is a positive semi-definite matrix for penalizing the terminal cost, $\bar{Q}\in \mathbb{R}^{n\times n}$ is the state penalizing matrix which is assumed to be positive semi-definite, and $\bar{R} \in \mathbb{R}^{m \times m}$ is a positive definite control penalizing matrix.

\subsection{Including Mode Sequence}
To solve the optimal switching problem, one needs to solve a two-level optimization \cite{Xu_parametrization}. In the upper level, switching times are assigned and in the lower level, the control policies to ensure the tracking are sought. To introduce the idea, consider a switched system with two subsystems and only one switching which happens at $t=t_1$. Also, let the mode sequence be $\{mode$ $1,$ $mode$ $2\}$. To make the switching time instant an independent parameter, let \cite{Xu_parametrization} 
\begin{equation}
t =
\begin{cases*}
t_0 + (t_1 - t_0) \hat{t} & if $0 \leq \hat{t}< 1$ \\
t_1 + (t_f - t_1) (\hat{t} - 1) & if $1\leq \hat{t} \leq 2$    
\end{cases*}
\label{eq_transformed_time}
\end{equation}  
%
using chain rule one can find $x'(\hat{t}) = \frac{dx}{d \hat{t} } =  \frac{dx}{dt} \frac{dt}{d\hat{t}}$ as 
%
\begin{equation}
x'(\hat{t})  =
\begin{cases*}
\Big(\bar{f}_1\big(x(\hat{t})\big) + \bar{g}_1\big(x(\hat{t})\big)u(\hat{t}) \Big) (t_1 - t_0) & if $0 \leq \hat{t}< 1$ \\
\Big(\bar{f}_2\big(x(\hat{t})\big) + \bar{g}_2\big(x(\hat{t})\big)u(\hat{t}) \Big) (t_f - t_1) & if $1\leq \hat{t} \leq 2$    
\end{cases*}
\label{eq_transformed_dynamics}
\end{equation}
Also, changing the independent variable in the cost function (\ref{cost_func_tracking_contin}) from $t$ to $\hat{t}$ leads
\begin{equation} 
\begin{split}
J(x_0) &\equiv J(t_1,x_0) = \big(x(2) - r(2)\big)^T S \big(x(2) - r(2)\big) + \int_{0}^{1} \frac{1}{2}\Big( \big(x(\hat{t}) - r(\hat{t})\big)^T \bar{Q} (t_1 - t_0) \big(x(\hat{t})- r(\hat{t})\big) + u(\hat{t})^T \bar{R}(t_1 - t_0) u(\hat{t}) \Big)d\hat{t}\\
& + \int_{1}^{2} \frac{1}{2} \Big( \big( x(\hat{t})- r(\hat{t})\big)^T \bar{Q} (t_f - t_1) \big(x(\hat{t})- r(\hat{t})\big) + u(\hat{t})^T \bar{R}(t_f - t_1) u(\hat{t}) \Big)d\hat{t}
\label{cost_switching_transformed}
\end{split}
\end{equation}
%
Letting $\delta \hat{t}$ as a small sampling time, by using Euler method one can discretize (\ref{eq_transformed_dynamics}) and (\ref{cost_switching_transformed}) as
\begin{equation}
x_{\hat{k}+1} =
\begin{cases*}
f_1\big(x_{\hat{k}} \big)  + g_1\big(x_{\hat{k}}\big) u_{\hat{k}}  & if $0 \leq \hat{k}< \frac{N'}{2}$ \\
f_2\big(x_{\hat{k}} \big)  + g_2\big(x_{\hat{k}}\big) u_{\hat{k}}  & if $\frac{N'}{2} \leq \hat{k} \leq N'$    
\end{cases*}
\label{eq_transformed_dynamics_for_lambda_DT}
\end{equation}
\begin{equation} 
\begin{split}
J(t_1, x_0)  &= \big(x_{N'} - r_{N'}\big)^T S \big(x_{N'} - r_{N'}\big) \\
& + \frac{1}{2}\sum_{\hat{k} = 1}^{N'/2} \Big( (x_{\hat{k}} - r_{\hat{k}} )^T \bar{Q} (t_1 - t_0) \delta \hat{t} (x_{\hat{k}} - r_{\hat{k}} ) +u_{\hat{k}}^T \bar{R}(t_1 - t_0) \delta \hat{t} u_{\hat{k}} \Big)\\
& + \frac{1}{2}\sum_{\hat{k} = N'/2+1}^{N'-1} \Big( (x_{\hat{k}} - r_{\hat{k}} )^T \bar{Q} (t_f - t_1) \delta \hat{t} (x_{\hat{k}} - r_{\hat{k}} ) + u_{\hat{k}}^T \bar{R}(t_f - t_1)\delta \hat{t} u_{\hat{k}} \Big)
\label{cost_switching_transformed_discrete}
\end{split}
\end{equation}
In (\ref{eq_transformed_dynamics_for_lambda_DT}), $f_1\big(x_{\hat{k}} \big) =  x_{\hat{k}} +  \bar{f}_1\big(x_{\hat{k}} \big) (t_1 - t_0) \delta \hat{t}$, $g_1\big(x_{\hat{k}} \big) =  \bar{g}_1\big(x_{\hat{k}} \big) (t_1 - t_0) \delta \hat{t}$, $f_2\big(x_{\hat{k}} \big) =  x_{\hat{k}} +  \bar{f}_2\big(x_{\hat{k}} \big) (t_f - t_1) \delta \hat{t}$, and $g_2\big(x_{\hat{k}} \big) = \bar{g}_2\big(x_{\hat{k}} \big) (t_f - t_1) \delta \hat{t}$. Also, in (\ref{eq_transformed_dynamics_for_lambda_DT}), $\hat{k} \in [1 ,N']$ is the discrete time index where $N' = \frac{number\hspace{3 pt} of \hspace{3 pt}switching + 1}{\delta \hat{t}}$ \cite{Heydary_fix_mode}. 
%
%
%
For finding the minimum cost-to-go from discrete time index $\hat{k}$ to $N'$(value function) one has\footnote{For $\hat{k}< N'/2$. Otherwise, $(t_1 - t_0)$ in (\ref{cost_to_go_1}) should be replaced by $(t_f - t_1)$. } 
\begin{equation} 
\begin{split}
V(t_1, x_{\hat{k}})  =& \min_{u(.)} \Big( \frac{1}{2}(x_{\hat{k}} - r_{\hat{k}} )^T \bar{Q} (t_1 - t_0) \delta \hat{t} (x_{\hat{k}} - r_{\hat{k}} ) +\frac{1}{2}u_{\hat{k}}^T \bar{R}(t_1 - t_0) \delta \hat{t} u_{\hat{k}}+  V(t_1, x_{\hat{k}+1}) \Big)
\label{cost_to_go_1}
\end{split}
\end{equation}
%
The optimal policy can be defined as
\begin{equation} 
\begin{split}
u^*_{{\hat{k}}}  = \argmin_{u(.)} \Big(& \frac{1}{2}(x_{\hat{k}} - r_{\hat{k}} )^T \bar{Q} (t_1 - t_0) \delta \hat{t} (x_{\hat{k}} - r_{\hat{k}} ) +\frac{1}{2}u_{\hat{k}}^T \bar{R}(t_1 - t_0) \delta \hat{t} u_{\hat{k}}+  V(t_1, x_{\hat{k}+1}) \Big)
\label{optimal_policy}
\end{split}
\end{equation}
%
Taking the gradient of the optimal value function, one can define the optimal costate as
\begin{equation} 
\lambda_{\hat{k}}(t_1, x_{\hat{k}}) =  
\begin{cases*}
{\mathscr{Q}}_1 + \frac{\partial x^*_{\hat{k}+1}}{\partial x_{\hat{k}}}\lambda_{\hat{k} + 1}(t_1, x^*_{\hat{k}+1})& if $0 \leq \hat{k}< \frac{N'}{2}$ \\
{\mathscr{Q}}_2 + \frac{\partial x^*_{\hat{k}+1}}{\partial x_{\hat{k}}}\lambda_{\hat{k}+1}(t_1, x^*_{\hat{k}+1}) & if $\frac{N'}{2} \leq \hat{k}\leq N'$ 
\label{costate_trans_swit}
\end{cases*}
\end{equation}
where 
\begin{equation}
\begin{split}
{\mathscr{Q}}_1 &= \bar{Q} (t_1 - t_0) \delta \hat{t} ( x_{\hat{k}}- r_{\hat{k}})\\
{\mathscr{Q}}_2 &= \bar{Q} (t_f - t_1) \delta \hat{t} ( x_{\hat{k}}- r_{\hat{k}})
\end{split}
\end{equation}
It is straight forward to see if optimal costates are known, one can directly calculate the optimal policy as 
\begin{equation} 
u^*_{\hat{k}} =  
\begin{cases*}
-(\bar{R} (t_1 - t_0) \delta \hat{t})^{-1} g_1^T(x_{\hat{k}})\lambda_{\hat{k}+1}(t_1, x^*_{\hat{k}+1}) & if $0 \leq \hat{k}< \frac{N'}{2}$ \\
-(\bar{R} (t_f - t_1) \delta \hat{t})^{-1} g_2^T(x_{\hat{k}})\lambda_{\hat{k}+1}(t_1, x^*_{\hat{k}+1}) & if $\frac{N'}{2} \leq \hat{k}\leq N'$ 
\label{policy_with_costates}
\end{cases*}
\end{equation}
In (\ref{costate_trans_swit}) and (\ref{policy_with_costates}), the super-script $*$ denotes the optimality. In the next subsection, a solution is presented to find the optimal costates. 

\subsection{ADP Solution} \label{sec_approach_1}

The backbone of the solution is training neural networks to approximate $\lambda_{\hat{k}+1}(t_1, x_{\hat{k}+1})$ from $(t_1,x_{\hat{k}})$. Based on Weierstrass Approximation Theorem \cite{Rudin}, linear-in-parameter neural networks with polynomial basis functions can uniformly \cite{HornikFeedforward} approximate continuous functions to a desired degree of precision in a compact set. Assuming that the value functions are continuously differentiable, one can use linear-in-parameter neural networks to approximate the costates. 
Hence, the approximate costates can be calculated as 
\begin{equation} 
\widehat{\lambda}_{\hat{k}+1}(t_1, x_{\hat{k}+1}) = \widehat{W}_{\hat{k}}^T \phi(t_1, x_{\hat{k}})
\label{approximate_costate}
\end{equation}
where $\widehat{W}_{\hat{k}} \in \mathbb{R}^{m_{\lambda}} \times \mathbb{R}^n$ is a tunable weight vector and $m_{\lambda}$ is the number of polynomial basis functions (neurons). The weight vectors are tuned through the training process backward in time. Once the costates are known, one finds the approximate optimal policy as 
\begin{equation} 
\widehat{u}_{\hat{k}}(t_1, x_{\hat{k}}) =  
\begin{cases*}
-R_1^{-1}g_1^T\big(x_{\hat{k}}\big) \widehat{\lambda}_{\hat{k}+1}(t_1, x_{\hat{k}+1})& if $0 \leq \hat{k}< \frac{N'}{2}$ \\
-R_2^{-1}g_2^T\big(x_{\hat{k}}\big) \widehat{\lambda}_{\hat{k}+1}(t_1, x_{\hat{k}+1})& if $\frac{N'}{2} \leq \hat{k}\leq N'$ 
\label{Policy_trans_swit}
\end{cases*}
\end{equation}
In (\ref{Policy_trans_swit}), $R_1= \bar{R} \delta \hat{t}(t_1 - t_0) $ and $R_2 = \bar{R} \delta \hat{t}(t_f - t_1)$. As mentioned before, for training, one can go backward in time and find the costates and save them for online control. 

Once the training concluded, one needs to find the optimal switching times from the costates for a selected initial condition $x_0 \in \Omega$. For this purpose, one can propagate the states along all possible switching times by using the costates and find the optimal cost to go for all possible switching time. Once done, one can choose the switching times which lead to the minimum cost.

\section{SIMULATION RESULTS}\label{sec_simulation_resutls}

To start the simulations, the values selected for the parameters of the WL model are given in Table \ref{table_para}. Some of these parameters are not introduced in this paper due to page constraints. Interested readers are directed to \cite{Nezhadali_compare_to_optimal} for more details. 
\begin{table}[t]
\renewcommand{\arraystretch}{1.3}
\caption{Parameters of WL model }
\label{table_para}
\centering
\begin{tabular}{| l | l |}
\hline
\bfseries Parameter &  \bfseries Description\\
\hline
$J_e$ =  0.43& engine inertia\\ \hline
$r_1$ =  2& lift arm dimensions \\ \hline
$r_2$ =  2 &lift arm dimensions \\ \hline
$\theta_1$ =  pi/6& bend angle of the lift arm\\ \hline
$n_{lc}$ =  2& number of lift cylinders\\ \hline
$A_{lc}$ =  0.0284& lift cylinder cross section area\\ \hline
$y_g$ =  2.13& height between body and lift arm hinge\\ \hline
$\eta_{lift}$ =  0.5 & lift system efficiency\\ \hline
$y_{off}$ =  0.5& boom hinge offset \\ \hline
$M_{tot}$ =  31330 &mass of WL and load\\ \hline
$\mu_{roll}$ =  0.03& rolling resistant factor \\ \hline
$F_{roll}$ =  $9.81 \mu_{roll} M_{tot}$& rolling resistance \\ \hline
$L$ =  3.7& WL wheel base\\ \hline
$\rho_f$ =  832& fuel density \\ \hline
$R_w$ =  0.3175 &wheel radius \\ \hline
$T_1$ = 5& constant for $P_{pump}$\\ \hline
$T_2$ =  -2.5& constant for $P_{pump}$\\ \hline
$\eta_{gb}$ =  0.9 &gearbox efficiency\\ \hline
$C_{st}$ =  $10^5$ &constant for steering power\\ \hline
$F_{buc}$ =  0.0981 $M_{tot}$ &  bucket and load weight\\ \hline
$F_{arm}$ =  $F_{buc}$ & arm weight\\ \hline
$I_{boom}$ =  1200& moment of inertia for the boom\\ \hline
\end{tabular}
\end{table}
Also, $W_{T_e} = [-0.154712845456646$, $0.555616949283938,$ $4.84689263976440,$ $0.450411638112411]^T$ in (\ref{eq_predicted_TE}) and the parameters in (\ref{eq_tau_p}) and (\ref{eq_p_stat}) are given as $a_1 = 3.07399910699804$, $a_2 = 0.467292644030092$, $a_3 = 0.467292644032403$, $\tau_1 = 5.94705195850280$, $\tau_2 = -0.703974251212264$. 

\begin{figure}[htbp]
\begin{center}
\includegraphics[scale = 0.9]{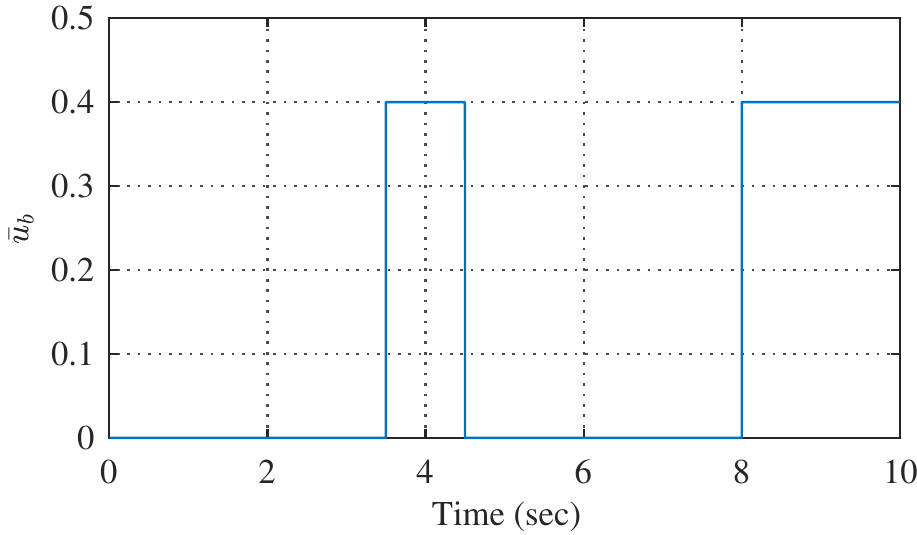}
\end{center}
\vspace{-5pt}	
\caption{Illustration of the normalized braking input $\bar{u}_b$. The maximum value for normalization was selected as $U_4 = 5 \times 10^5$. 
} 
\label{Fig_WL_response_u_b}
\vspace{-10pt}
\end{figure}

\begin{figure}[htbp]
\begin{center}
\includegraphics[scale = 0.9]{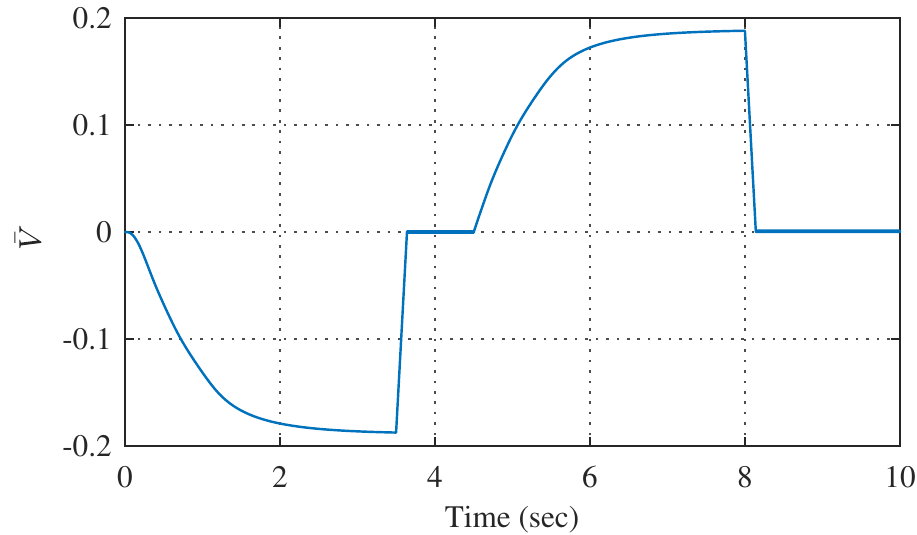}
\end{center}
\vspace{-5pt}	
\caption{Illustration of the normalized velocity of the WL in the presented SLC. 
} 
\label{Fig_WL_response_V}
\vspace{-10pt}
\end{figure}

\begin{figure}[htbp]
\begin{center}
\includegraphics[scale = 0.9]{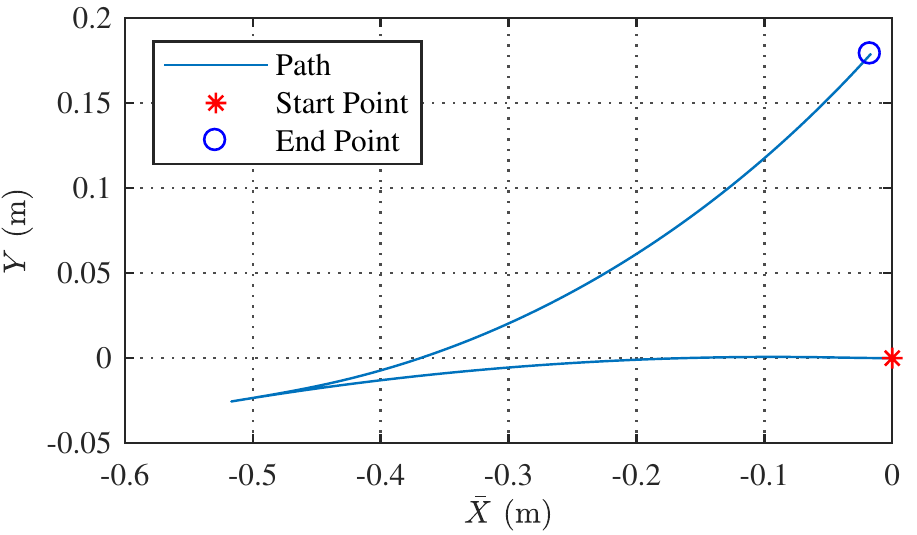}
\end{center}
\vspace{5pt}	
\caption{Illustration of the path of wheel loader with the selected inputs. The start and end points are highlighted in the figure. The wheel loader goes a V-shape path which simulates an SLC. } 
\label{Fig_WL_response_X_Y}
\vspace{-10pt}
\end{figure}

In order to see the performance of the model, some selected control inputs are given to the WL, and the performance of the WL is evaluated. The mode sequence in this example is $\{$backwarding, stopping, forwarding, stopping$\}$. All the switching time instants and the continuous control inputs to the wheel loader are selected manually to simulate a short loading cycle. 
The braking input, vehicle velocity and the path are shown in Figures \ref{Fig_WL_response_u_b}, \ref{Fig_WL_response_V}, and \ref{Fig_WL_response_X_Y}, respectively.

In order to apply the ADP method to find the optimal switching time and control, one assumes that the reference signal is a known function of time. For this example, the reference signal is selected as $\dot{r}(t) = [0,0,0,0,\sin(\pi t), \sin(\pi t), 0,0,0,0,0]$. The start and end times are selected as $t_0 = 0$, and $t_f = 3$. In this example, the mode sequence is selected as $\{backwarding,\hspace{3pt}stopping \}$. In other words, half of an SLC is considered. Since there is only one switching in the system, by choosing $\delta \hat{t} = 0.001$, one can calculate $N' = 2000$. The state and control penalizing matrices are selected as $S = \diag([0,0,0,0,10^4,10^4,10^4,0,0,0,0])$, $\bar{Q} = \diag([0,0,0,0,10^4,10^4,0,0,0,0,0])$, and $\bar{R} = \diag([1,1,1,1])/ \delta \hat{t}$ where $\diag([a,b])$ is a diagonal matrix with values of $a$ and $b$ on the main diagonal and zero elsewhere. In choosing $S$ matrix for terminal state penalizing, one notes that penalization of the velocity is enforced as it is important to bring the vehicle to rest at the final time. 

\begin{figure}[htbp]
\begin{center}
\includegraphics[scale = 0.3]{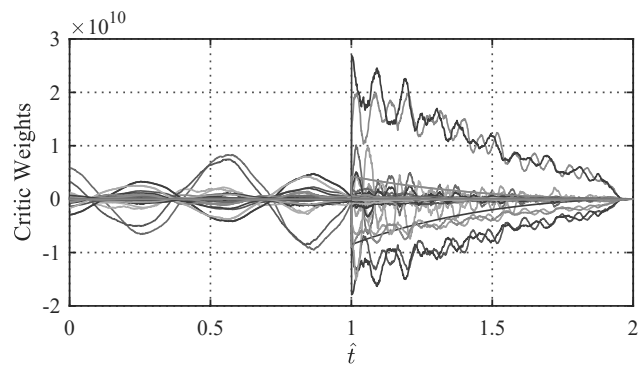}
\end{center}
\vspace{-30pt}	
\caption{The history of the neural network to approximate the costates. } 
\label{fig_weights_his}
\vspace{-10pt}
\end{figure}

In order to start the training, one needs to select a good neural network to capture the dynamics of the costates. Mostly, such selection is performed by trial and error. 
After selecting a relatively rich set of basis functions, the training process was performed. The history of the neural network weights is shown in Figure \ref{fig_weights_his}. As one can see from Figure \ref{fig_weights_his}, the history of the weights shows a jump at $\hat{t} = 1$ which is the switching time. 

\begin{figure}[htbp]
\begin{center}
\includegraphics[scale = 0.9]{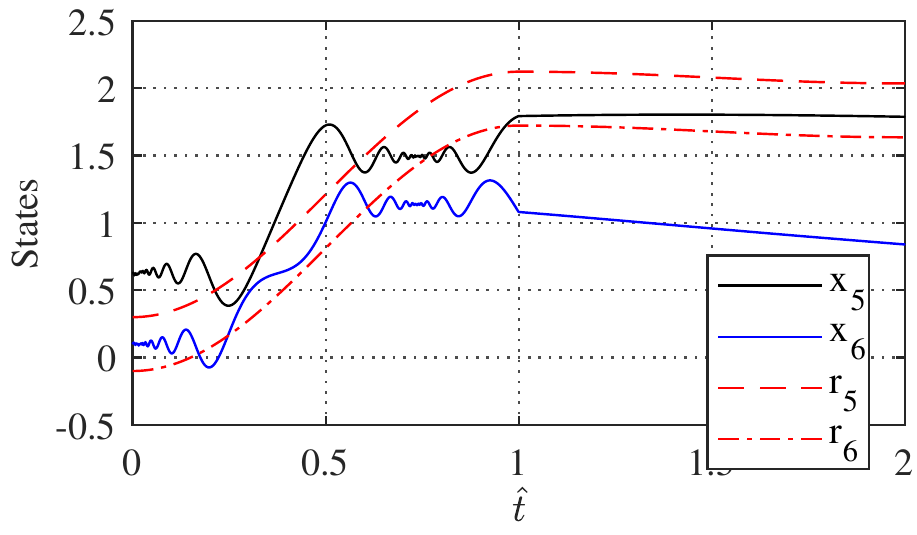}
\end{center}
\vspace{5pt}	
\caption{State history in the closed loop system. The optimal switching time was found as $t_1 = 2.86$. } 
\label{States_Not_good}
\vspace{-10pt}
\end{figure}

When the training process concluded, a random initial condition was selected in the domain of training. The initial condition for the reference signal was selected as $r_0 = [0,0,0,0,0.3,-0.1,0,0,0,0,0]^T$ and 
the optimal switching time was found as $t_1 = 2.86$.
Afterward, the optimal costates were used to propagate the states from the previously selected random initial condition. The history of the states is shown in Figure \ref{States_Not_good}. The poor tracking is the result of approximation errors in capturing the dynamics of the costates and it is ongoing research with our research group to solve this problem. 
 
\section{CONCLUSION} \label{sec_conclusion}
An approximate dynamic programming solution is used for optimal control of a wheel loader. For this purpose, the dynamics of the wheel loader are modeled as a switched system with controlled subsystems and a fixed mode sequence. Some simulation results are provided to show the effectiveness of the solution. 


\bibliography{References_WL}
\bibliographystyle{IEEEtran}

\end{document}